\documentclass[aps, prd, twocolumn, showpacs, floatfix, letterpaper, nofootinbib, amsmath, amssymb, superscriptaddress]{revtex4-1}

\usepackage[utf8]{inputenc}  
\usepackage[T1]{fontenc}     
\usepackage{amsmath,amssymb,lmodern} 
\usepackage{graphicx}
\usepackage{diagbox}
\usepackage{hyperref}
\usepackage{color}
\usepackage{lipsum}
\usepackage{bm}
\usepackage{url}
\usepackage{relsize}
\usepackage{ulem}

\usepackage{xcolor}
\usepackage{color}
\definecolor{azul}{rgb}{0.0, 0.58, 0.71} 
\definecolor{vermelho}{rgb}{0.8, 0.0, 0.0}
\definecolor{verde}{rgb}{0.0, 0.26, 0.15}
\definecolor{amarelo}{rgb}{0.8, 0.5, 0.2}
\definecolor{vermelhoescuro}{rgb}{0.59, 0.0, 0.09}
\definecolor{reddish}{rgb}{0.65, 0.2, 0.2}
\definecolor{darkgreen}{rgb}{0.2,0.7,0.3}
\definecolor{darkblue}{rgb}{0.3,0.40,0.48}
\definecolor{gray}{rgb}{.8,.8,.8}

\hypersetup{,
    unicode=false,          % non-Latin characters in Acrobat bookmarks
    pdftoolbar=true,        % show Acrobat toolbar?
    pdfmenubar=true,        % show Acrobat menu?
    pdffitwindow=true,     % window fit to page when opened
    pdfstartview={FitH},    % fits the width of the page to the window
    pdftitle={Multiple anisotropic bounces},    % title
    pdfauthor={A. Bacalhau, P. Peter and S. Vitenti},     % author
    pdfnewwindow=true,      % links in new window
    colorlinks=true,       % false: boxed links; true: colored links
    linkcolor=reddish,          % color of internal links
    citecolor=darkblue,        % color of links to bibliography
    filecolor=magenta,      % color of file links
    urlcolor=darkblue,           % color of external links
    linktocpage=true
}

\usepackage[disable]{todonotes}

\def\setR{\mathbb{R}}
\def\setC{\mathbb{C}}

\newcommand{\bV}{b_\mathrm{v}}

\newcommand{\ex}{\mathrm{e}}
\newcommand{\dd}{\mathrm{d}}
\newcommand{\GN}{G_{_\mathrm{N}}}

\newcommand{\Mp}{M_{_\mathrm{P}}}

\renewcommand{\(}{\left(}
\renewcommand{\)}{\right)}
\renewcommand{\[}{\left[}
\renewcommand{\]}{\right]}

\begin{document}

\title{Anisotropic multiple bounce models}

\author{Anna Paula Bacalhau}
\email{anna@cbpf.br}
\affiliation{ICRA -- Centro Brasileiro de Pesquisas 
F\'isicas, Xavier Sigaud, 150, Urca, Rio de Janeiro, Brazil}

\author{Patrick Peter}
\email{peter@iap.fr}
\affiliation{Institut d'Astrophysique de Paris (${\cal
G}\setR\varepsilon\setC{\cal O}$), UMR 7095, and Institut Lagrange de
Paris \\ Universit\'e Pierre \& Marie Curie (Paris 6) et CNRS, Sorbonne
Universit\'es \\ 98 bis boulevard  Arago, 75014 Paris, France}
  
\author{Sandro D. P. Vitenti}
\email{sandro.vitenti@uclouvain.be}
\affiliation{Centre for Cosmology, Particle Physics and Phenomenology,
Institute of Mathematics and Physics, Louvain University, 2 Chemin
du Cyclotron, 1348 Louvain-la-Neuve, Belgium}
\affiliation{ICRA -- Centro Brasileiro de Pesquisas 
F\'isicas, Xavier Sigaud, 150, Urca, Rio de Janeiro, Brazil}

\begin{abstract}
We analyze the Galileon ghost condensate implementation of a bouncing
cosmological model in the presence of a non negligible anisotropic
stress. We exhibit its structure, which we find to be far richer than
previously thought. In particular, even restricting attention to a single
set of underlying microscopic parameters, we obtain, numerically, many
qualitatively different regimes: depending on the initial conditions on
the scalar field leading the dynamics of the universe, the contraction
phase can evolve directly towards a singularity, avoid it by bouncing
once, or even bounce many times before settling into an ever-expanding
phase. We clarify the behavior of the anisotropies in these various
situations.
\end{abstract}


\date{\today}
\maketitle

\section{Introduction} \label{Intro}

Observational cosmology \cite{Ade:2015lrj} mostly favors a primordial
phase of single field slow-roll inflation \cite{Lemoine:2008zz,
Martin:2013nzq} (see Ref.~\cite{Ijjas:2015hcc} for an alternative
analysis of these data however). Moreover, inflation remains the most
widely accepted paradigm permitting to solve the usual big bang puzzles
of horizon, flatness and entropy \cite{PPJPU2013}, leading to the
currently observed homogeneous and isotropic universe. It also provides a
natural means of producing linear perturbations with an almost
scale-invariant spectrum. Non inflationary scenarios, however unfavored,
have also been proposed, which yield an almost scale-invariant spectra
for the scalar modes \cite{Battefeld:2014uga, Brandenberger:2016vhg}. The
next step will perhaps come from observation of the tensor modes, whose
signal could be used to discriminate various models, thus giving the
necessary tools to discard or to accept non singular and non inflationary
bouncing models.

The very first interest of a non-singular cosmology is of course that it
avoids the singularity inherent to ever-expanding scenarios, thereby
allowing to increase the size of the horizon as much as needed: with a
long contraction phase followed by a bounce connecting to the presently
expanding universe, any region can have been in causal contact with any
other. Moreover, providing the contraction is decelerated for a
sufficiently long time, the universe reaches the bounce in an almost flat
condition. In short, bouncing models can also solve many of the
above mentioned standard big bang puzzles in ways that differ from the
inflationary solutions \cite{Peter:2008qz}.

The so-called ``matter''-bounce
\cite{Finelli:2001sr,Brandenberger:2012zb} belongs to this particular
class of models which may confront observations. During a contraction
dominated by a field with negligible effective pressure $p \ll \rho$,
the long wavelength scalar perturbations, originating in vacuum state,
reach the bounce phase with a scale invariant  spectrum
\cite{Finelli:2001sr}. Another way to realize such a spectrum was
proposed in Ref. \cite{Khoury:2001wf}: in this so-called ekpyrotic 5
dimensional model, the motion of 4 dimensional branes yields an
effective 4 dimensional theory experiencing a contraction followed by
an expansion phase in the Einstein frame; this model however needs to
pass through a singular phase spoiling predictability \cite{Martin2002,
Martin:2002ar}. In terms of the effective
Friedmann-Lema\^\i{}tre-Robertson-Walker (FLRW) space-time, the
dynamics can be mimicked by assuming the stress-energy tensor to be
that of a scalar field $\phi$ with a negative potential describing the
relative motion of the branes. In the proposal of
\cite{Finelli:2001sr}, the exponential potential for the scalar field
results in $w \equiv p/\rho \sim 0$, for a specific choice of
parameters.

An important issue possibly plaguing any bouncing cosmological model,
including in particular the  matter bounce proposal, is the excessive
growth of any initially small anisotropy deviation during contraction;
this is usually referred to as the Belinsky, Khalatnikov and Lifshitz
(BKL) \cite{Belinsky:1970ew} instability. Since the anisotropic stress
goes with the scale factor $a(t)$ as $a^{-6}$ (i.e. it can be seen as a
fluid with effective equation of state $w_\sigma=1$), it can eventually
dominate over the other fluid energy densities if the bounce is deep
enough and/or the initial anisotropy is too large. This scenario
assumes that the other fluids have equations of state $w<1$
($w_\mathrm{rad}=\frac13$ or $w_\mathrm{mat}\sim 0$), during the
contraction phase. Note however that, if the anisotropy is initially
very small, for instance resulting from initial quantum vacuum
fluctuations, the bounce can take place before the anisotropy dominates
over the other matter components~\cite{Vitenti2012, Pinto-Neto2014}.

A way to solve the anisotropy problem in a contracting universe
followed by a non-singular bounce, without merely setting its initial
value to being vanishingly small,  was suggested in
Ref.~\cite{Buchbinder:2007ad}. In this model, the potential was also
chosen to be exponential so as to give the scalar field fluid an
effective equation of state $w_\phi\gg 1$. As a result, this field
dominates over the anisotropic stress during the entire contraction
epoch, hence preventing any growth of the anisotropy compared to the
other components. By the end of the contraction phase, a second scalar
field takes over in a ghost condensate state, i.e., one for which the
kinetic term may develop a non vanishing minimum inducing the canonical
kinetic term in the Lagrangian to change sign for a finite amount of
time. This leads to an effective equation of state $w_\phi <-1$, which
drives the scale factor evolution to a halt: the universe goes through
a bouncing phase. The condensate is responsible not only for the
non-singular transition between contraction and expansion, but it also
corrects the wavelength dependence of the scalar modes
\cite{ArkaniHamed:2003uy}: this is required because the ekpyrotic
contraction yields a non scale-invariant spectrum for the
perturbations.

Such a model has to face two problems. The first comes from the scalar
field in the ghost condensate state: in order that it does not
interfere with the background dynamics during the ekpyrotic
contraction, it should be sufficiently diluted so as to dominate only
during the final stages of contraction. The second issue concerns the
long wavelength scalar perturbations, which may grow unstable after
exiting the ekpyrotic phase \cite{Xue:2011nw}, leading to a spectrum
very different from the observed quasi scale invariant one.

To address these issues, in yet another version of this new ekpyrotic
model, the ghost condensate is obtained via a Galileon term that
couples the scalar field with the metric~\cite{Qiu:2013eoa}. By means
of two different functions of the scalar field, namely a negative
potential $V(\phi)$ controlling the ekpyrotic phase, and a non-standard
kinetic coupling $g(\phi)$ controlling the ghost condensate, it was
then argued that the anisotropy growth is suppressed and the
non-singular bounce is achieved even in the presence of small
anisotropic deviation \cite{Cai:2012va, Cai:2013vm}. A curvaton
mechanism \cite{Enqvist:2001zp, Lyth:2001nq, Moroi:2001ct} is then
invoked to finally produce scale invariant perturbations in the
expansion phase.

Present calculations of the perturbations in models such as those
discussed above have been done assuming an FLRW perturbed metric, under
the assumption that the anisotropic stress can be made negligible for
the relevant scales. On the other hand, if this assumption is not
strictly valid and the background space-time is in fact Bianchi I, at
least in some range of times, then it was shown \cite{Pereira:2007yy}
that the scalar, vector and tensor modes evolve in a coupled way
already at first order. Even for an inflationary phase, this is known
to yield possible effects in the resulting spectrum
\cite{Pitrou:2008gk}, and it is only natural to expect a similar
conclusion to hold in a contracting universe model. This could
drastically modify any prediction for the final perturbation spectrum
produced in such a model. Before addressing this question however, it
is necessary to discuss the dynamics of the background itself.

The present work aims at exploring the evolution stemming from the
theory proposed in~\cite{Cai:2013vm}; as it happens, it is much richer
than previously anticipated. The non singular models studied so far
were for the most part based on one non-singular FLRW bounce. We show
here that the highly non-linear features of the dynamical equations
lead to a variety of unforeseen different scenarios. We exhibit four
examples for which the underlying microscopic parameters are chosen as
in Ref.~\cite{Cai:2013vm}: they lead respectively to a singularity,
one, two or even three bouncing stages depending on the chosen initial
conditions. Our purpose is to exemplify these possibilities in order to
open up and possibly constrain the relevant dynamical phase space of
the acceptable backgrounds from which one will have to subsequently
study the perturbations, to be eventually compared with the
observational data.

The paper is organized as follows: in the following Section, we review
the model of Ref.~\cite{Cai:2013vm}, expand the relevant equations of
motion in the Bianchi I case and set the dynamical system to be solved
numerically. Section~\ref{Sec:NumSol} is devoted to a presentation of
our numerical solutions for different background behaviors, pointing
out the main phenomenology behind the different dynamics.
Section~\ref{Sec:Disc} returns to the basic equations and discusses the
role, influence and evolution of the anisotropy in multiple bounce
scenarios, together with a discussion on the expected effects of
changing the model parameters. Section~\ref{Sec:Concl} summarizes our
findings and offers some concluding remarks and expectations on the
power spectrum and its properties in these models.

In what follows, we work in units such that $\hbar = c = 1$ and the
reduced Planck mass is $\Mp = 1 / \sqrt{8 \pi \GN}$. The metric
signature is $(+, -, -, -)$. Throughout the paper, the scale factor is
normalized to unity at the first bouncing point.

\section{General equations} \label{Sec:GenEq}

Our starting point is to assume a Galileon scalar field $\phi$ minimally
coupled to Einstein gravity, i.e.
\begin{equation}
\mathcal{S} = \int \dd^4 \sqrt{-g} \( \frac12 \Mp^2 R +\mathcal{L}\),
\label{thTOT}
\end{equation}
where the scalar field Lagrangian is taken to be
\begin{equation}
\mathcal{L}\left[ \phi \left(x\right)\right] = K(\phi, X) + G(\phi, X) \Box \phi ,
\label{Lagrangian}
\end{equation}
$K$ and $G$ being functions of the field itself and its canonical kinetic term
\begin{equation}
 X \equiv \frac12 \partial_\mu \phi \partial^\mu \phi,
\end{equation}
and $\Box\phi \equiv g^{\mu\nu} \nabla_\mu \nabla_\nu \phi$ where
$\nabla_\nu$ represents the torsion-free covariant derivative compatible
with the metric $g_{\mu\nu}$.

Variations of $\mathcal{L}$ yields the relevant energy momentum tensor
\begin{eqnarray}
T^{\phi}_{\mu\nu} &=&
(-K+2XG_{,\phi}+G_{,X}\nabla_\sigma{X}\nabla^\sigma\phi)g_{\mu\nu}
\nonumber\\
&& + (K_{,X}+G_{,X}\Box\phi-2G_{,\phi})\nabla_\mu\phi\nabla_\nu\phi
\nonumber\\
&& - G_{,X}(\nabla_\mu{X}\nabla_\nu\phi+\nabla_\nu{X}\nabla_\mu\phi),
\label{energystress}
\end{eqnarray}
where the notations $F_{,\phi}$ and $F_{,X}$ stands for derivatives of
whatever $F$ stands for with respect to $\phi$ and $X$, respectively.

Following Ref. \cite{Cai:2013vm}, we choose
\begin{equation}
K(\phi, X) = \Mp^2 \left[1-g(\phi) \right]X + \beta X^2 - V(\phi),
\label{Kessence}
\end{equation}
with the positive-definite parameter $\beta$ ensuring the kinetic term
to be bounded from below at high energy scales and we assume the scalar
field $\phi$ is dimensionless, hence the Planck mass coefficient on the
first term. The arbitrary functions in \eqref{Kessence} must be such as
to render an ekpyrotic contraction phase together with a non singular
ghost condensate dominated bounce possible. As explained in
\cite{Cai:2013vm}, an acceptable choice is provided by
\begin{equation}
g(\phi) =
\frac{2g_0}{\ex^{-\sqrt{\frac{2}{p}}\phi}+\ex^{b_g\sqrt{\frac{2}{p}}\phi}},
\label{gphi}
\end{equation}
with $g_0>1$, $p>0$ and $b_g$ dimensionless constants, while the
potential can be taken as
\begin{equation}
V(\phi) =
-\frac{2V_0}{\ex^{-\sqrt{\frac{2}{q}}\phi}+\ex^{\bV\sqrt{\frac{2}{q}}\phi}},
\label{Vphi}
\end{equation}
where $V_0>0$ is constant with dimension of $({\rm mass})^4$ and two
other dimensionless constants $q$ and $\bV$. This negative-definite
potential reduces to the exponential form of the ekpyrotic scenario
\cite{Finelli:2001sr} for large values of $\phi$. Finally, the function
$G(\phi, X)$ is of the Galileon type \cite{Deffayet:2009wt}, again chosen
in agreement with \cite{Cai:2013vm} as $G(X) = \gamma X$, with $\gamma$
is a positive dimensionless constant.

With the matter content fixed, the system is complete once we give the
relevant geometrical symmetries. This we do by assuming a flat,
homogeneous and anisotropic universe, whose dynamics is described by a
Bianchi I metric, namely
\begin{equation}
\dd s^2 = \dd t^2 - a^2(t) \sum_{i}\ex^{2\theta_i(t)} \dd x^i\dd x^i,
\label{Bianchimetric}
\end{equation}
the average scale factor $a(t)$ permitting to define a mean Hubble rate
through $H \equiv \dot{a}/a$, the ``dot'' denoting time derivative with
respect to cosmic time $t$.

The equation of motion of the scalar field $\phi$ is derived from the
Lagrangian \eqref{Lagrangian} and can be cast in the form of a modified
Klein-Gordon equation, namely
\begin{equation}
\mathcal{P} \ddot\phi + \mathcal{D} \dot\phi +V_{,\phi} = 0,
\label{eom}
\end{equation}
where the functions $\mathcal{P}$ and $\mathcal{D}$ depend on both the
scalar field itself and the geometry; they are respectively given by
\begin{equation}
\mathcal{P} = (1-g)\Mp ^2 +6\gamma H\dot\phi +3\beta\dot\phi^2
+\frac{3\gamma^2}{2\Mp ^2}\dot\phi^4,
\label{Pterm}
\end{equation}
and
\begin{align}
\mathcal{D} =& 3(1-g)\Mp ^2H +\left(
9\gamma{H}^2-\frac{1}{2}\Mp^2g_{,\phi}\right) \dot\phi
+3\beta{H}\dot\phi^2 \nonumber\\ &
-\frac{3}{2}(1-g)\gamma\dot\phi^3 -\frac{9\gamma^2H\dot\phi^4}{2\Mp^2}
-\frac{3\beta\gamma\dot\phi^5}{2\Mp ^2} % \nonumber\\ &&
-\frac{3}{2}\gamma\sum_i\dot\theta_i^2\dot\phi.
\label{Fterm}
\end{align}
The parameters of the model are $g_0$, $V_0$, $b_g$, $\bV$, $p$, $q$,
$\beta$, $\gamma$ all real, positive and assumed non vanishing. Without
lack of generality, we set $\Mp\to 1$ for the rest of this work.

Defining the shear
\begin{equation}
\sigma^2 = \sum_i\dot\theta_i^2,
\label{def_sigm}
\end{equation}
the Friedmann equations follow from the stress-energy tensor
\eqref{energystress}; they are
\begin{equation}
H^2 = \frac{\rho_{\phi}}{3} + \frac{\sigma^2}{6},
\label{Friedmann1} 
\end{equation}
for the constraint, and
\begin{equation}
\dot{H} = -\frac{\rho_{\phi} + p_{\phi}}{2} - \frac{1}{2}\sigma^2.
\label{Friedmann2}
\end{equation}
In \eqref{Friedmann1} and \eqref{Friedmann2}, the energy
density $\rho_\phi$ and pressure $p_\phi $ of the scalar field are
given by
\begin{eqnarray}
\rho_\phi &=& \frac{1}{2} (1-g)\dot\phi^2 +\frac{3}{4}\beta\dot\phi^4 
+3\gamma H\dot\phi^3 +V(\phi), \label{rho} \\
p_\phi &=& \frac{1}{2} (1-g)\dot\phi^2 +\frac{1}{4}\beta\dot\phi^4 
-\gamma\dot\phi^2\ddot\phi -V(\phi). \label{pressure}
\end{eqnarray}
Finally, as discussed in the introduction, the shear evolves as
\begin{equation}
\sigma^2 = \sigma^2_\mathrm{ini}\left(\frac{a_\mathrm{ini}}{a}\right)^6,
\label{shear_dyn}
\end{equation}
i.e., as a stiff-matter fluid, where the subscript ``ini'' denotes an
arbitrary initial time. For future convenience, we shall refer to the
quantity
\begin{equation}
\rho_{\sigma} \equiv \frac{\sigma^2}{2} = p_\sigma
\label{den_sigma}
\end{equation}
as the energy density and pressure associated to the anisotropy.

\section{Numerical solutions} \label{Sec:NumSol}

The dynamical equations presented in the last section can be recast
into a system of first order differential equations, namely
\begin{align}
\dot{\phi} &= \varphi, \\
\dot{\varphi} &= - \frac{\mathcal{D} \varphi}{\mathcal{P}} - \frac{V_{,\phi}}{\mathcal{P}}, \\
\dot{H} &= - \frac{\rho_{\phi} + p_{\phi}}{2} - \frac{\sigma^2_\mathrm{ini}}{2}\left(\frac{a_\mathrm{ini}}{a}\right)^6, \\
\dot{a} &= a H,
\end{align}
where we have introduced a new variable $\varphi$ to reduce the system
order and used Eqs.~\eqref{eom}, \eqref{Friedmann2},
\eqref{shear_dyn} and the definition of the mean Hubble rate $H$. We
assume the underlying parameters are those already chosen in
\cite{Cai:2013vm}, so the numerical solutions presented below will be
comparable with this previous work. We have
\begin{align*}
 V_0   &=  10^{-7} \Mp^4, &    g_0 &= 1.1,\\
 \bV   &= 5,              &    b_g &= 0.5,\\
 p     &= 0.01,           &      q &= 0.1,\\
 \beta &= 5,              & \gamma &= 10^{-3}.
\end{align*}
The initial conditions are given by the set 
\begin{equation}\label{key}
\theta = (\phi_\mathrm{ini},\; \varphi_\mathrm{ini}) \ \ \ \hbox{and}
\ \ \ \sigma^2_\mathrm{ini} = 5\times 10^{-12},
\end{equation}
with $\varphi_\mathrm{ini}$ chosen in such a way that the kinetic
contribution $\propto \varphi_\mathrm{ini}^2$ be comparable to the
shear contribution at the initial time (recall we fixed
$a_\mathrm{ini}=1$), while $H$ is given by the
constraint~\eqref{Friedmann1}, namely
\begin{equation}
H_\mathrm{ini} = - \sqrt{\frac{{\rho_{\phi}}_\mathrm{ini}}{3} + 
\frac{1}{6}\sigma^2_\mathrm{ini}},
\end{equation}
where ${\rho_{\phi}}_\mathrm{ini}$ is obtained with Eq. \eqref{rho}
evaluated at $\phi_\mathrm{ini}$ and $\varphi_\mathrm{ini}$. Finally,
note that we have omitted the scale factor $a$ since it enters
explicitly only in the expression for the shear in Eq.~\eqref{shear_dyn}
through the combination $\sigma_\mathrm{ini} a_\mathrm{ini}^3$:
without loss of generality, one can renormalize the initial shear to
account for the initial value of the scale factor, which can thus be
chosen as $a_\mathrm{ini} = 1$ for simplicity.

Reference~\cite{Cai:2013vm} considered the presence of a matter
component, $p \ll \rho $, assumed to produce the initially
scale-invariant spectrum. Here, we want to focus on the bounce itself, or
the behavior of the scale factor in general when the universe is
dominated by the scalar field. This means we begin our analysis at a time
for which we assume the dust fluid contribution has already turned
negligible, having been overcome by the other components  when we set our
initial conditions. In other words, for $a < a_{\mathrm{ini}}$ (we set
initial conditions in a contracting epoch), the matter fluid is
negligible and we shall accordingly forget it altogether.

In the numerical solutions presented below in Figs.~\ref{Fig:1B_H_a}
through \ref{Fig:rho_sigma}, the time $t$ is expressed in units of
$10^4 \Mp^{-1}$ and the Hubble rate $H$ in units of $10^{-4} \Mp$. In
order to compare the solutions with the same reference point, we always
set the initial time to $t_{\mathrm{ini}} = 0$. The estimated  absolute
error in the calculations shown are of order $\mathcal{O}(10^{-10})$
during the contraction and expansion epochs, and $\mathcal{O}(10^{-7})$
during the bounce phase. Since we are interested in the potential
effects of a remaining sub-dominant anisotropy during the bounce, we
consider in what follows initial conditions such that the effective
equation of state (EoS) is not very large at the beginning, i.e., we
are assuming that only a weak ekpyrotic phase, where the EoS is only
slightly above one, has taken place in our scenario before we set our
initial conditions.

\subsection{One bounce scenario} \label{Sec:1b}

\begin{figure}[ht]
\includegraphics[width=0.5\textwidth]{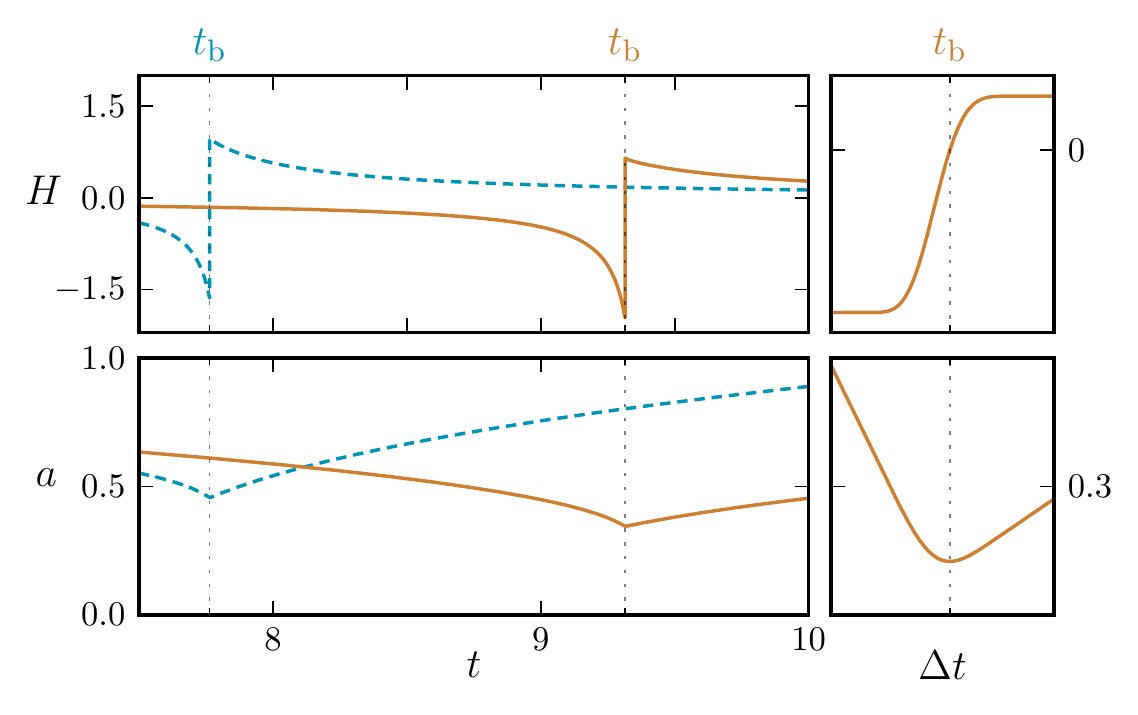}
\caption{Time evolution of the Hubble constant $H$ (top left) and scale
factor $a$ (bottom left) for $\varphi_\mathrm{ini} = 8 \times 10^{-6}$
and two different values of $\phi_{\mathrm{ini}}$: $\phi_{\mathrm{ini}}
= -3$ (full brown) and $\phi_{\mathrm{ini}} = -2.5$ (dashed blue). The
bounce times are marked as $t_{\mathrm{b}}$. The discontinuity is only
apparent and a mere consequence of the fact that the relevant time
scale is extremely short for the fast bounce that takes place in this
theory: the right panels show the details of this actually smooth
transition (shown only for $\phi_\mathrm{ini} = -3.5$) over the much
smaller time interval of $\Delta t = 10^{-4}$ around the bounce time
$\textcolor{amarelo}{t_\mathrm{b}}$. Although not shown explicitly on
later plots, all the following curves are in fact smooth on the
relevant scales as we did check for all cases.}
\label{Fig:1B_H_a}
\end{figure}

The single bounce scenario is the most widely discussed background
evolution for bouncing cosmologies. The background evolves dominated by
the scalar field during contraction, passes through the ghost
condensate phase, makes a single non singular bounce and enters an ever
lasting expansion phase afterwards, as exemplified in
Fig.~\ref{Fig:1B_H_a}. These numerical solutions were obtained for
$\phi_{\mathrm{ini},1} = -2.5$ and $\phi_{\mathrm{ini},2} = -3.0$, with
$\varphi_\mathrm{ini} = 8 \times 10^{-6}$ in both cases. As noted
earlier, the initial shear value is close to the kinetic term
$\varphi^2 \sim \times 10^{-11}$, and is subsequently diminished (in
comparison to $\rho$) during the ekpyrotic phase.

\begin{figure}[ht]
\includegraphics[width=0.5\textwidth]{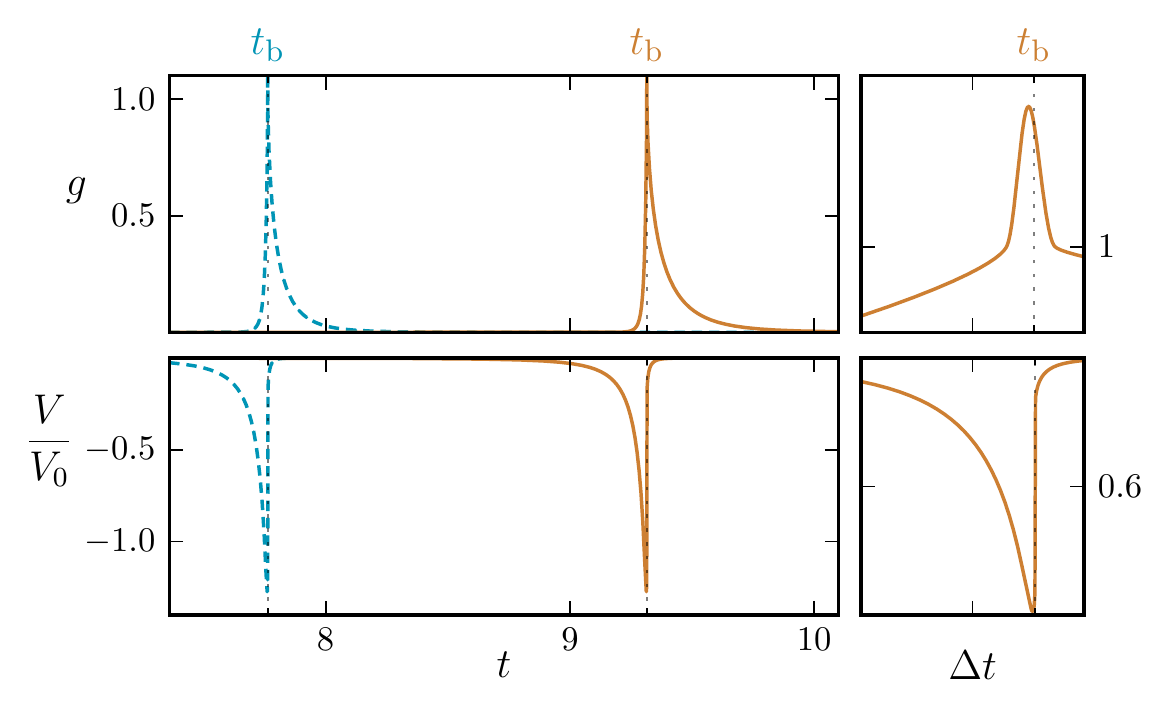}
\caption{Time development of the kinetic function $g\[\phi\(t\)\]$ (top
left) and potential $V\[\phi\(t\)\]/V_0$ (bottom left), with the same
convention as Fig.~\ref{Fig:1B_H_a}. The ghost condensate phase begins
as soon as $g(\phi) \geq 1$. The right panel shows how smooth the
transition goes when looked at on shorter timescales.}
\label{Fig:1B_g_V}
\end{figure}

\begin{figure*}[ht]
\includegraphics[width=\textwidth]{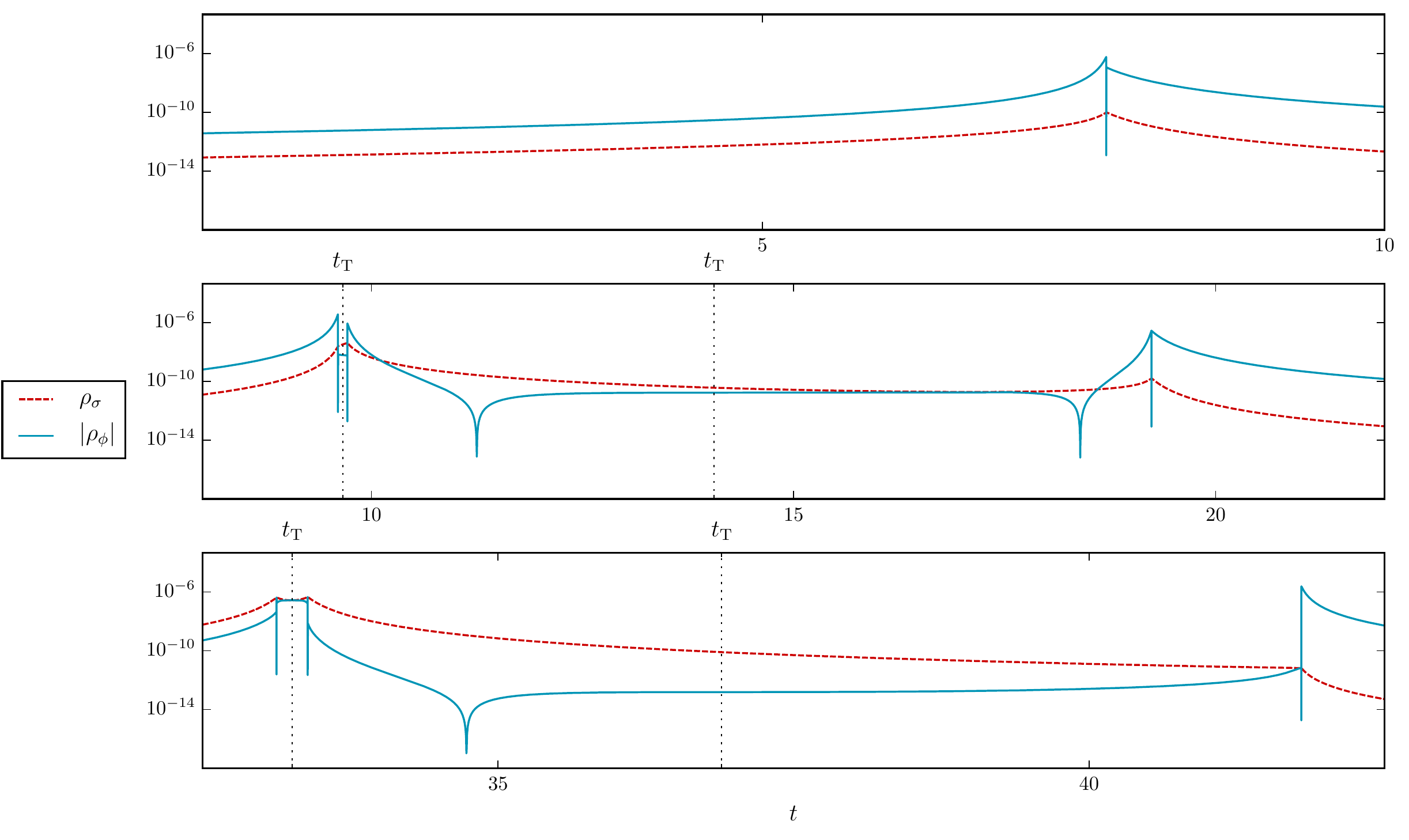}
\caption{Comparative evolution of the energy densities for the
anisotropy, $\rho_{\sigma}$ (red dashed) and the scalar field,
$\rho_{\phi}$ (blue full) for the initial conditions
$\{ \phi_\mathrm{ini} = -2.5,\ \varphi_\mathrm{ini} = 8 \times
10^{-6}\}$ (top, single bounce), $\{ \phi_\mathrm{ini} = -3.5,\ 
\varphi_\mathrm{ini} = 8 \times 10^{-6}\}$ (middle, two bounces) and
$\{ \phi_\mathrm{ini} = 1.9,\  \varphi_\mathrm{ini} = - 10^{-6}\}$
(bottom, three bounces). The initial anisotropic stress for all
the plots is $\sigma^2_\mathrm{ini} = 5\times10^{-12}$. The indicated
$t_\mathrm{T}$ are the turning points at which the scalar field
goes through the maximum of $g(\phi)$. }
\label{Fig:rhos}
\end{figure*}

The ghost condensate and ekpyrotic phases are presented in
Fig.~\ref{Fig:1B_g_V} where the time development of the kinetic term
coefficient $g$ and the potential $V$ are presented. Before the bounce
takes place, the scalar field is driven by the potential which becomes
very negative all through the ekpyrotic phase, until $g$ takes over, at
which point the bounce occurs. Fig.~\ref{Fig:rhos} shows, for this case
and the following (with more than one bounce taking place), the time
evolution of the energy contained in the scalar field and in the shear.
The top panel is for the case at hand: the difference between
$\rho_{\phi}$ and $\rho_{\sigma}$ is entirely due to $V(\phi)$ in this
case, and as expected, the shear contribution decreases with respect to
that of the field.

Reducing the shear is what the ekpyrotic phase is made for. Indeed,
with the potential \eqref{Vphi}, there exists an attractor solution with
EoS for the scalar field $w_\phi$
\begin{equation}
 w_{\phi} \approx -2 + \frac{2}{3q},
 \label{w_eff}
\end{equation}
while on the other hand, Eq.~\eqref{shear_dyn} implies that the EoS of
the shear is $w_{\sigma} = 1$. For small values of $q$, as the one we
are interested in and have chosen for the numerical calculations, the
shear can never dominate during contraction. Although we did not start
on the attractor \eqref{w_eff}, we obtained that behavior for two
different choices of initial conditions for $\phi_\mathrm{ini}$. The
more negative $\phi_{\mathrm{ini}}$, the longer the contraction phase,
because the scalar field begins farther away from the ghost condensate
state that permits the bounce. There is a degeneracy in the initial
condition space, since one could achieve a similar behavior by changing
$\varphi_\mathrm{ini}$, an initially small velocity for the field
leading to a longer contraction phase as it takes more time to reach
the ghost condensate phase.

At first sight, one is tempted to conclude from the previous discussion
that $\phi_\mathrm{ini}$ or $\varphi_\mathrm{ini}$ could be chosen as
small as one wishes in order to yield a longer contraction phase and
varying the bounce characteristic features. As it turns out, this is
not the case at all: as we show in the following section, changing the
initial conditions produces drastically different solutions involving
more than one bounce.

\begin{figure}[ht]
\includegraphics[width=0.5\textwidth]{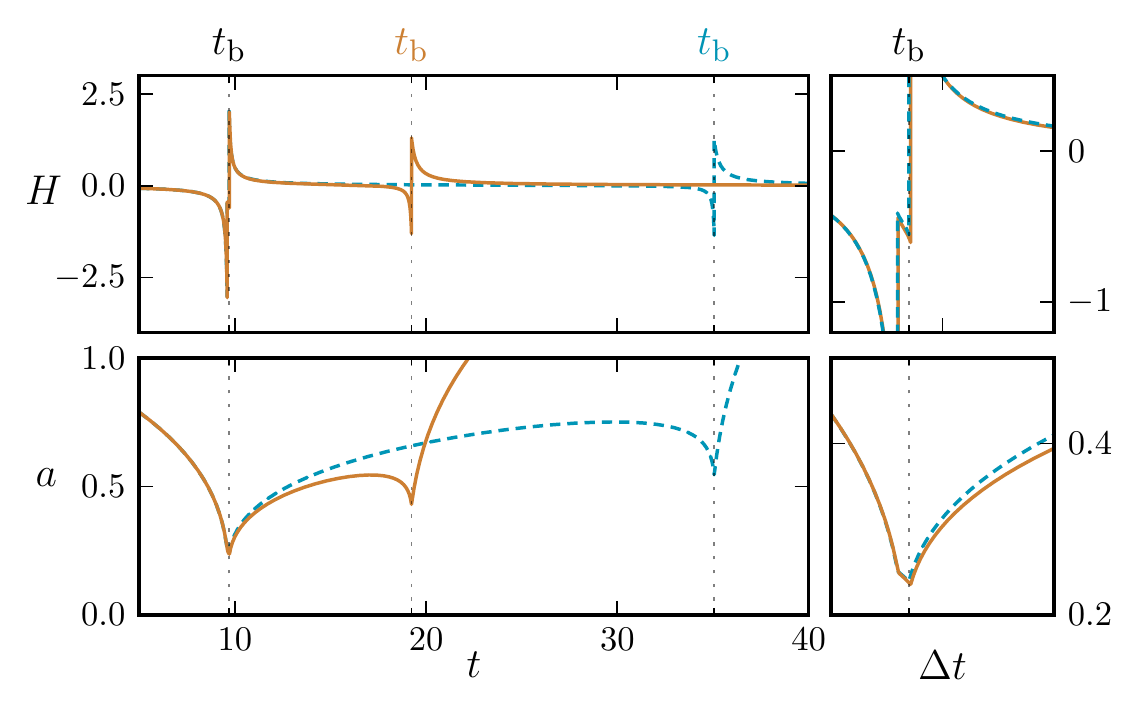}
\caption{Evolution of the Hubble parameter $H$ (top left) and the scale
factor $a$ (bottom left) for the two different initial conditions:
$\phi_{\mathrm{ini}} = -3.5$ (full yellow) and $\phi_{\mathrm{ini}} =
-3.49$ (blue dashed). The bounces are marked as $t_\mathrm{b}$. The
first bounce of the two solutions are indistinguishable on the figure
(numerically extremely close), but the solutions then drift away and
bifurcate, yielding a second bounce at very different times, first for
$\phi_{\mathrm{ini}} = -3.5$, then for $\phi_{\mathrm{ini}} = -3.49$.
This indicates an extreme sensibility in the initial conditions that
has never been discussed in such a context. The plots on the right
detail  what happens during the first time the system goes through the
ghost condensate phase, with a time scales of the plot taken as $\Delta t
\approx 3$ around $\textcolor{amarelo}{t_\mathrm{b}}$.}
\label{Fig:2B_H_a}
\end{figure}

\subsection{Two bounce case} \label{Sec:2b}

\begin{figure}[ht]
\includegraphics[width=0.5\textwidth]{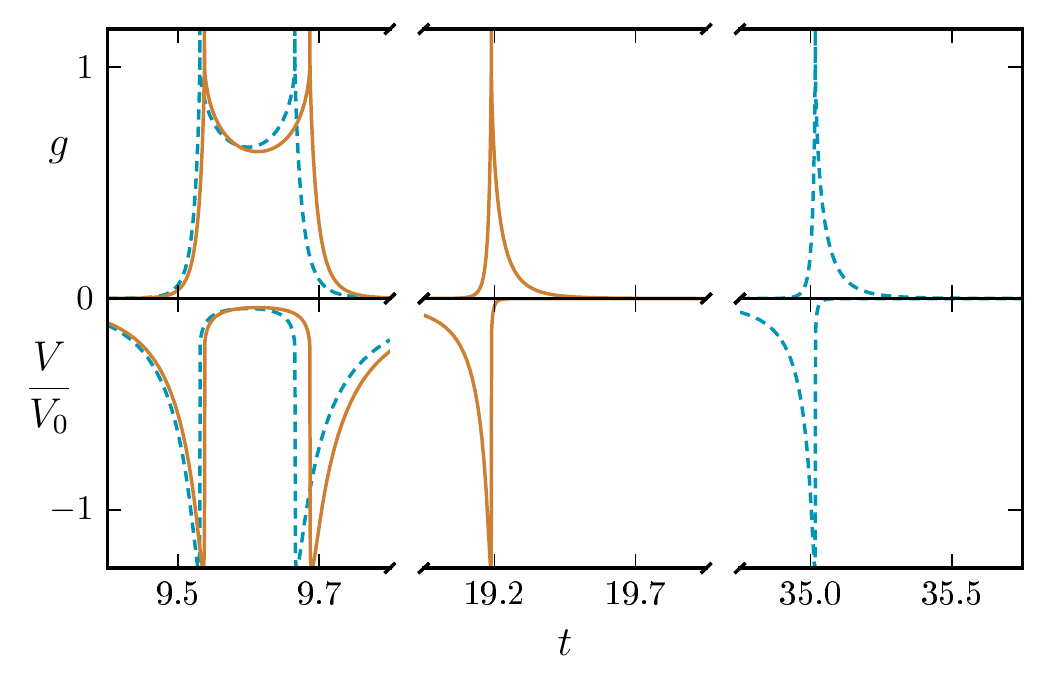}
\caption{Time developments of $g(\phi)$ (top) and the potential
$V(\phi)$ (bottom). The two solutions are for $\phi_\mathrm{ini} =
-3.5$ (full yellow), $\phi_\mathrm{ini} = -3.49$ (blue dashed), both
with $\varphi_\mathrm{ini} = 8 \times 10^{-6}$. As in Fig.
\ref{Fig:1B_g_V} the peaks only appear discontinuous but
in actuality are smooth.}
\label{Fig:2B_g_V}
\end{figure}

Figure~\ref{Fig:2B_H_a} illustrates what happens if one keeps decreasing
$\phi_\mathrm{ini}$, trying to trigger a longer contraction phase: one
reaches a region in parameter space in which the Universe instead
experiences two bounces. The universe contracts, bounces, expands again,
passes through a maximum, starts contracting again and moves towards a
second bounce, from which it finally expands forever. For that to happen,
the scalar field must go twice through the ghost condensate phase, a
possibility which was always assumed hard to achieve, whereas in fact, we
found it actually goes through this phase three times (see Fig.
\ref{Fig:2B_g_V}) even though only two bounces took place. 

This evolution is exemplified by $\varphi_\mathrm{ini} = 8 \times
10^{-6}$ and the two initial field conditions
$\phi_{\mathrm{ini},1}=-3.49$ and $\phi_{\mathrm{ini},2}= -3.50$, whose
subsequent time development is depicted in Figs.~\ref{Fig:2B_H_a} and
\ref{Fig:2B_g_V}.

The behavior we find here is due to the existence of a turning point
for $\phi$, marked as $t_\mathrm{T}$ in Fig.~\ref{Fig:rhos}. At this
point, the scalar field passes through the first ghost condensate phase
while still contracting. It eventually returns and goes back to pass
through the top of the potential $g(\phi)$ another time. Then, the
universe bounces.

In Fig.~\ref{Fig:2B_g_V}, we show that after the first bounce took
place, the expansion phase is again dominated by the ekpyrotic
potential $V(\phi)$. As we mentioned before, during the ekpyrotic
phase, the effective EoS of the scalar field is built to be larger than
that of the anisotropy. This means that, during contraction, the scalar
field dominates for small values of $a$, but conversely also that
during expansion, the anisotropy becomes more and more important. This
is illustrated in Fig.~\ref{Fig:rhos} where the shear domination after
the first bounce is clearly visible.

With the expansion dominated by the anisotropy, $\phi$ reaches a second
turning point, while $H$ became negative again. This is the beginning
of the second contraction phase that will eventually drive $\phi$ into
the ekpyrotic phase again (see the third peak of Fig.
\ref{Fig:2B_g_V}), thereby reducing the shear contribution again. When
the scalar field again reaches the peak of $g(\phi)$, (third ghost
condensate phase), this triggers the bounce in an even more isotropic
state.

From that example, one can envisage two possible scenarios. Without the
first turning point, the Universe would have gone through a ghost
condensate phase without triggering a bounce and a singularity would
have ensued. It is often stated that one of the most dangerous effect
that can prevent a bounce from taking place is the uncontrolled growth
of anisotropy. We found that the scalar field initial conditions are
also important in order to ensure the bounce can occur. Below we also
argue that in fact, it is thanks to the existing anisotropy that the
universe does not plunge straightforwardly into a singularity. The
second scenario is when conditions are such as to avoid the second
turning point altogether. In that case, the last expansion epoch begins
anisotropic: the ekpyrotic contraction, although controlling the
relative shear decay, is not sufficient as the multiple bounces
subsequently spoil its effect. A phase of ekpyrotic contraction is thus
not necessarily enough to guarantee that the resulting universe, after
the bounce, expands isotropically, the scalar field initial conditions
playing a crucial role in the overall evolution of the universe.

\subsection{Three bounces} \label{Sec:3b}

Our final example is rather counter intuitive. It begins with an
anisotropic contraction phase not leading to a BKL instability and
resulting into a final expansion phase even more isotropic than the
previous cases (see Fig.~\ref{Fig:rhos}). To produce this scenario, we
tune the value of $\phi_\mathrm{ini}$, chosen positive, keeping the
amount of initial anisotropy as before, $\sigma^2_\mathrm{ini} =
5\times 10^{-12}$, and we set $\varphi_\mathrm{ini} = -10^{-6}$,
together with the two field values $\phi_{\mathrm{ini},1}=1.9$ and
$\phi_{\mathrm{ini},2}=1.9001$, noting that since the initial field
time derivative is smaller, the anisotropy is  initially larger than
the kinetic term $\varphi^2 = 10^{-12}$.

\begin{figure}[ht]
\includegraphics[width=0.5\textwidth]{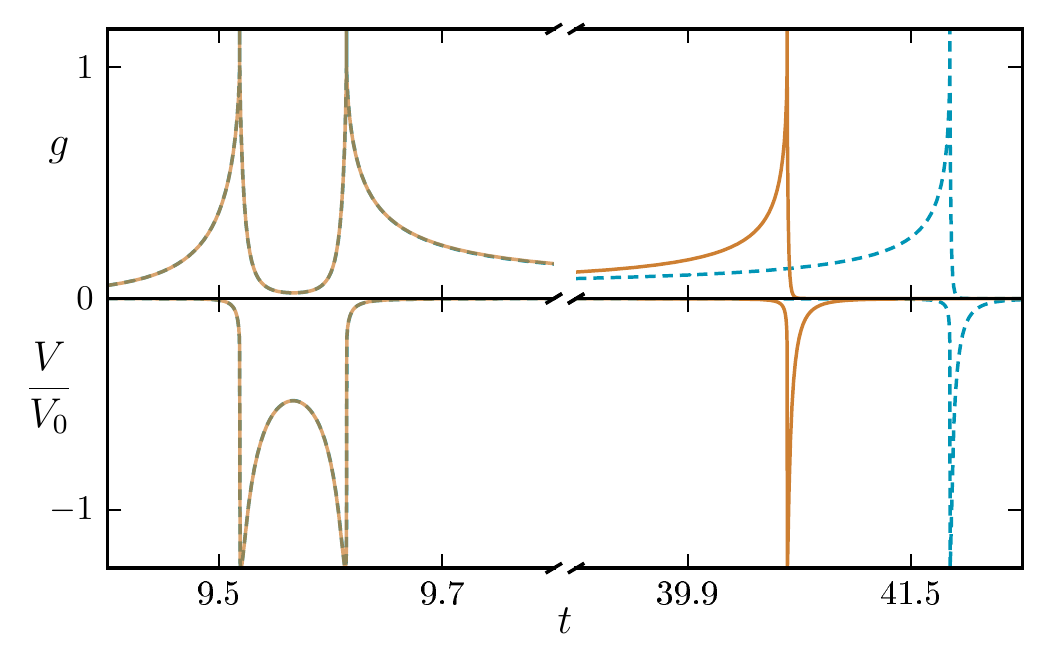}
\caption{Time developments of $g(\phi)$ (top) and the potential
$V(\phi)$ (bottom). The three peaks leads to the three bounces of
Fig.~\ref{Fig:3B_H_a} with initial conditions given by $\phi_\mathrm{ini} =
1.9001$ (full yellow), and $\phi_\mathrm{ini} = 1.900$ (blue dashed).
The fine-tuning required on $\phi_\mathrm{ini}$ reflects the fact that
it is extremely difficult to obtain a final isotropically expanding
state when beginning with a shear dominated contracting universe. In
fact, almost any other initial condition leads to a singularity.}
\label{Fig:3B_g_V}
\end{figure}

The usual ekpyrotic approach consists in beginning with the ekpyrotic
phase so as to lower, dissolve really, the relative shear contribution
immediately, during the initial contraction, thereby solving the
anisotropy problem. The case here is completely different, as we start
with $\phi_\mathrm{ini} >0$ and $\varphi_\mathrm{ini} < 0$ so that the
scalar field starts evolving from the right hand side of the potential
$V(\phi)$ and of $g(\phi)$. This means that, contrary to the cases
discussed above, we do not begin the evolution of the universe with the
ekpyrotic phase: this phase only happens after the first ghost
condensate peak, as shown in Fig.~\ref{Fig:3B_g_V}.

As in the two bounce case of Sec.~\ref{Sec:2b}, the existence of a
turning point is mandatory for the observed behavior. Otherwise, the
universe merely collapses into a singularity.

\begin{figure}[ht]
\includegraphics[width=0.5\textwidth]{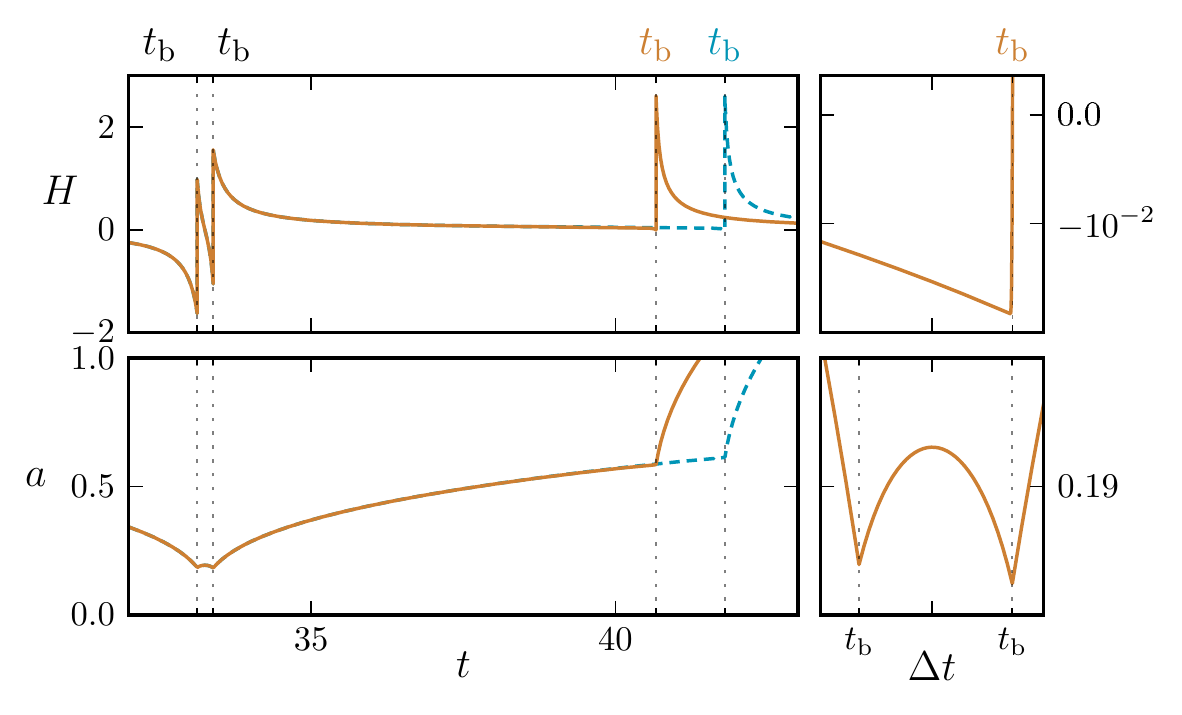}
\caption{Evolution of the Hubble constant $H$ (top left) and the scale
factor $a$ (bottom left) for the same initial conditions as in
Fig.~\ref{Fig:3B_g_V}. The other initial conditions for both cases are
$\varphi_\mathrm{ini} = - 10^{-6}$ and $\sigma^2_{\mathrm{ini}}=
5\times10^{-12}$. The first two bounces happen at roughly the same time for
both initial conditions, and the solutions then drift away as in the
previous example before reaching the third bounce. The top right panels
emphasizes the smoothness of the evolution of $H$ around the third
bounce in the case $\phi_{\mathrm{ini}} = 1.9001$ (the other has a
similar shape). It turns out the Hubble scale becomes slightly
negative only, and for a very limited amount of time, indicating a very
short contraction phase. The bottom left panel details the first two
bounces for the case $\phi_{\mathrm{ini}} = 1.9001$. The time scale of
the plots are $\Delta t \approx 10^{-3}$ around the third bounce,
$\textcolor{amarelo}{t_\mathrm{b}}$ (top right panel) and $\Delta t
\approx 10^{-1}$ around the first two bounces, indicated by
$t_\mathrm{b}$ (bottom right). Enlarging more the time scale on this
latter plot shows the bounces are, again, smooth and only appear
discontinuous because of the time scales used to represent them.}
\label{Fig:3B_H_a}
\end{figure}

The presence of three ghost condensate phases, i.e., the peaks of
$g(\phi)$ in Fig.~\ref{Fig:3B_g_V}, leads to the three bounces of
Fig.~\ref{Fig:3B_H_a}. The first contraction, containing no ekpyrotic
phase, is completely dominated by the anisotropy (Fig.~\ref{Fig:rhos}).
After the first bounce, the universe expands ekpyrotically as it
reaches the first peak of $V(\phi)$, Fig. \ref{Fig:3B_g_V}. During this
ekpyrotic expansion, $\phi$ reaches a turning point and $H$ changes
sign, initiating the second contraction.

After the second contraction, the universe once again goes through the
ghost condensate phase and another bounce occurs. The ensuing expansion
is still anisotropic, until the scalar field reaches another
turning point, at which point the universe begins contracting for the
third time while $\phi$ climbs back up in $g(\phi)$. During this third
contraction, which is not ekpyrotic-like, the scalar field energy
contribution appears to grow faster than the anisotropy, as shown
in~Fig. \ref{Fig:rhos}. The scenario ends after $\phi$ crosses the last
peak of $g(\phi)$, and the universe bounces for the third time.

As can be seen in Fig.~\ref{Fig:3B_H_a}, the third contraction is a
very short phase with a minimum Hubble scale of $H_\mathrm{min} \approx
10^{-2}$ before the third bounce. Because the contraction was shorter
than the expansion, the anisotropy is more diluted. At the same time,
$\phi$ starts to grow faster than the anisotropy. This is a very
unexpected behavior. As we can see in Fig. \ref{Fig:3B_g_V}, there is
no ekpyrotic potential contribution before the third bounce to render
the effective EoS of the scalar field larger than that of the
anisotropy.

The final stage of the process described above is the third bounce
itself, at which point the scalar field overcomes the anisotropy,
leading the universe to the required isotropic expansion. Even though
the expansion in dominated by the scalar field in the ekpyrotic phase
(Fig. \ref{Fig:3B_g_V}), the difference between the energy densities is
large enough that the anisotropy does not end up dominating.

\subsection{Singular solutions}

\begin{figure}[ht]
\includegraphics[width=0.5\textwidth]{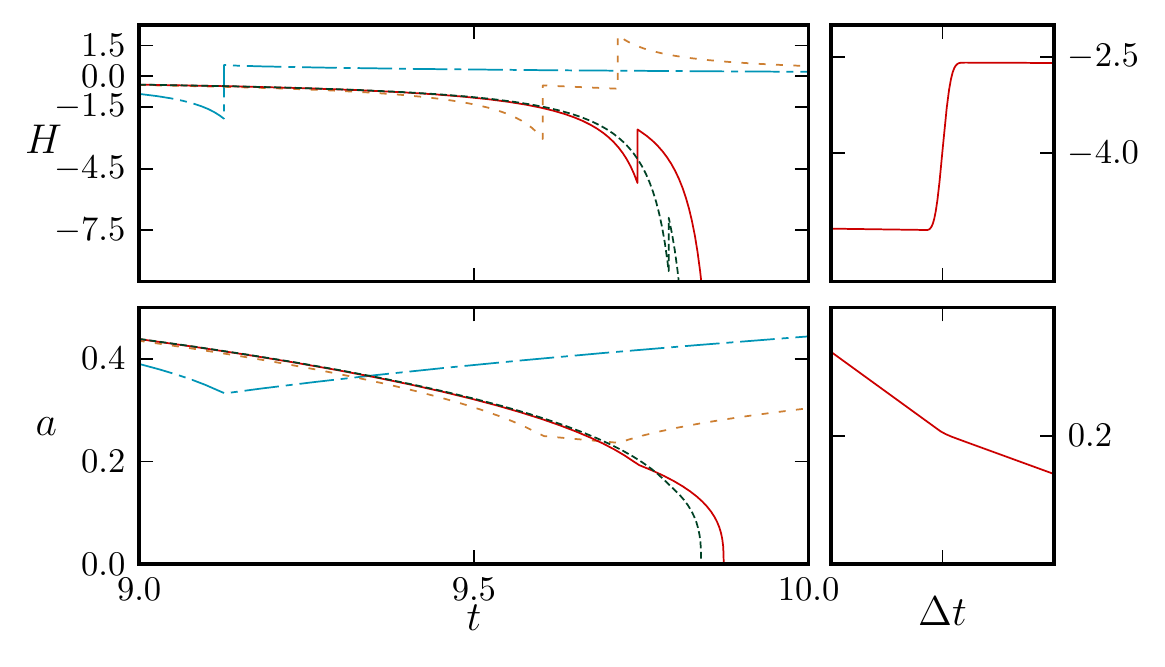}
\caption{Time evolution of the Hubble constant, $H$ (top left), and the
scale factor, $a$ (bottom left), for $\dot{\phi}_\mathrm{ini} = - 8
\times 10^{-6}$ and $\sigma^2_{\mathrm{ini}}  = 5 \times 10^ {-12}$, with four
different initial conditions on $\phi_\mathrm{ini}$ leading
respectively to one bounce (blue dot-dashed line, $\phi_\mathrm{ini} =
-2.5$), two bounces (yellow small dashed line, $\phi_\mathrm{ini} =
-3.5$) and singular solutions (red full line, $\phi_\mathrm{ini} =
-4.0$ and green long dashed line, $\phi_\mathrm{ini} = -4.5$). The
right panel details what happens at the point where the dynamics would
lead to a bounce in a regular solution: the system goes through the
ghost condensate, but for an insufficient amount of time, so that even
though $H$ increases (top right), changing the slope of $a$ (bottom
right), it remains negative, leading ultimately to an unavoidable
singularity. The time scale for the right panel plots is $\Delta t =
10^{-4}$ around $t = 9.7$.}
\label{Fig:S_H_a}
\end{figure}

Despite the presence of an ekpyrotic phase and a ghost condensate regime,
the existence of a bouncing solution is not guaranteed. In Fig.
\ref{Fig:S_H_a}, we show a sequence of solutions for different values of
$\phi_\mathrm{ini}$, assuming in all cases $\varphi_\mathrm{ini} = 8
\times 10^{-6}$ and $\sigma^2_\mathrm{ini}= 5\times10^{-12}$, some solutions
being regular and bouncing, other contracting endlessly to a singularity,
for initial values of the scalar field not too far away from one another.
The list of initial conditions used here is $\phi_{\mathrm{ini},1} = -2.5$,
$\phi_{\mathrm{ini},1} = -3.5$, $\phi_{\mathrm{ini},1} = -4.0$, and 
$\phi_{\mathrm{ini},1} = -4.5$.

This last case leads us to conclude that the more negative
$\phi_\mathrm{ini}$, the longer the contraction phase and the larger
the anisotropy when the system reaches the ghost condensate state.
Fig.~\ref{Fig:S_H_a} shows the transitions from one bounce, two bounces
and no bounce solutions while decreasing $\phi_\mathrm{ini}$. As it turns out,
the singular solution is not the limit of a single bounce case, but rather
a two-bounce situation in which the second bounce is failed, the
Hubble rate suddenly increasing while the scalar field passes through
the ghost condensate phase, but not enough to render it positive, so the
ghost condensate epoch terminates in a still contracting phase, and
the universe has subsequently no chance to return to expansion. 

\section{Discussion} \label{Sec:Disc}

The main feature enabling the universe described by our model to bounce
a few times is the possibility of one or more turning points,
making the scalar field  climb the potentials more than once. Let us
now examine their properties and the consequences they can have on the
evolution of the universe.

As discussed earlier, a turning point at time $t_\mathrm{T}$ is
characterized by $\varphi(t_\mathrm{T}) = 0$ and
$\dot{\varphi}(t_\mathrm{T}) < 0$, if $\phi(t_\mathrm{T})$ is a local
maximum or $\dot{\varphi}(t_\mathrm{T}) > 0$, if $\phi(t_\mathrm{T})$
is a local minimum. This solution should necessarily satisfy the
Friedmann equation \eqref{Friedmann1}. Substituting $\varphi =
\dot{\phi}=0$ in Eq.~\eqref{Friedmann1} and defining
\begin{equation}
y \equiv \ex^{\sqrt{\frac{2}{q}} \phi},
\label{def_y}
\end{equation}
the Friedmann constraint reads:
\begin{equation}
y^{-1} + y^{\bV} + \frac{2}{3} V_0\( H^2 -
\displaystyle\frac{\sigma^2}{6} \)^{-1} = 0.
\label{poly}
\end{equation}
Demanding that there exists a turning point means that there should be
at least one root to the above equation.

Let us define $f(y)$ from Eq.~\eqref{poly}, namely
\begin{equation}
f(y) =  y^{-1} + y^{\bV} + \frac{2}{3} V_0\( H^2 -
\displaystyle\frac{\sigma^2}{6} \)^{-1}.
\label{def_poly}
\end{equation}
The range of variation of this function follows from that of $\phi$, so
that $y \rightarrow 0$ implies $\phi \rightarrow - \infty$,  and
$y\rightarrow \infty$ leads to $\phi \rightarrow  \infty$. One can
easily check that
\begin{equation}
\lim_{y \rightarrow 0} f(y) = \infty \ \ \ \ \hbox{and} \ \ \ \
\lim_{y \rightarrow \infty} f(y) = \infty,
\end{equation}
so that if there exists $\bar{y}$ such that $f(\bar{y})<0$, then, by
virtue of Bolzano's theorem on intermediate values for continuous
functions, one is guaranteed that $f(y)$ possesses at least two roots,
which we call respectively $\phi^{*}_1 \in \big]- \infty , \sqrt{q/2} \ln
\bar{y}\big]$ and $\phi^{*}_2 \in \big[ \sqrt{q/2} \ln \bar{y},
\infty\big[$, eligible as turning points.

The condition $f(\bar{y})<0$ is only possible provided the condition
\begin{equation}
H^2 < \frac{\sigma^2}{6}. \label{cond_sigma}
\end{equation}
holds. This last expression shows that the anisotropy plays a non
trivial part in the existence of the turning point, enabling $f(y)$ to
have a root. For an isotropic universe, the shear by definition
vanishes, $\sigma^2 = 0$ and the system can only bounce once. By
continuity, for very small values of the initial anisotropy, the
initial condition on $\phi$ dictates whether it is possible that
Eq.~\eqref{cond_sigma} is satisfied. The higher the shear, the more
likely one encounters a regime during which Eq.~\eqref{cond_sigma} is
valid during the evolution and thus, the more likely the existence of
turning points. Contrary to the common lore, a high value of the
primordial anisotropy may therefore not necessarily spoil the bouncing
scenario, or even the resulting isotropic expansion.

\begin{figure}[ht]
\includegraphics[width=0.5\textwidth]{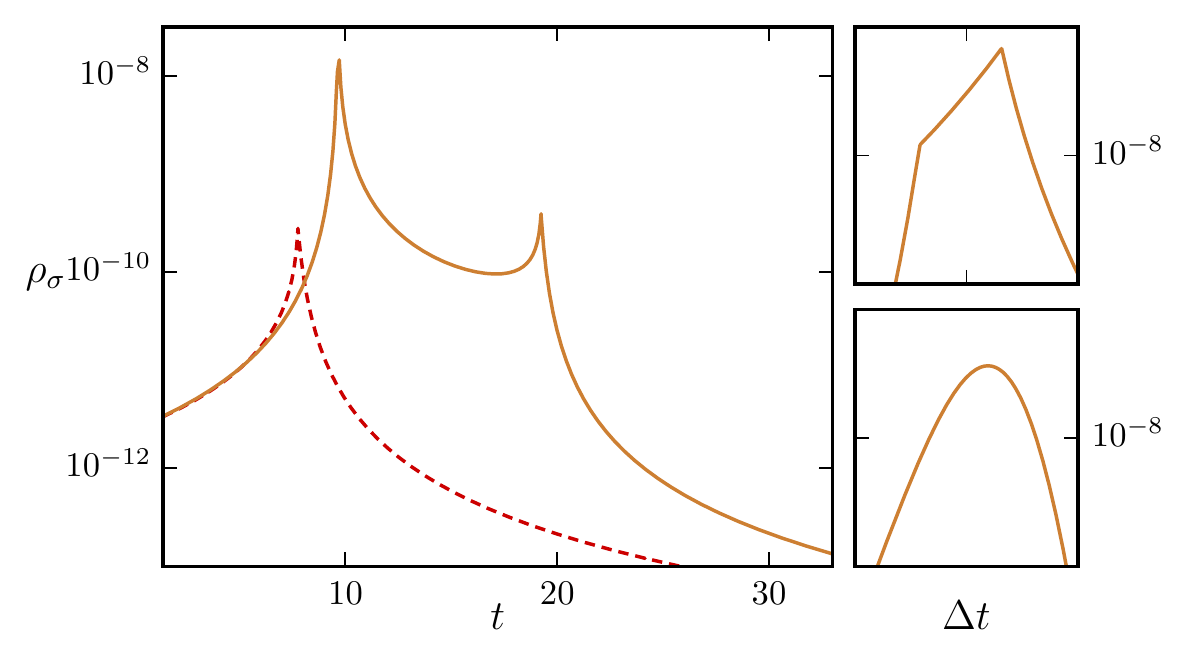}
\caption{Evolution of the effective energy density for the anisotropic stress, 
$\rho_{\sigma}$ (left) for $\varphi_\mathrm{ini} = 8 \times 10^{-6}$ and 
$\sigma^2_{\mathrm{ini}} = 5\times10^{-12}$ with $\phi_\mathrm{ini} = -2.5$ (red 
dashed line, single bounce) and $\phi_\mathrm{ini} = -3.5$ (full yellow line, 
two bounce case), previously discussed in Sections \ref{Sec:1b} and \ref{Sec:2b} 
respectively. The top right panels shows a detail of the first peak of the 
anisotropy energy density in the two bounces case. We can see the effect of the 
first turning point and the ekpyrotic expansion in the increase of the 
anisotropy before the first bounce, highest peak. The bottom right panel is a 
zoom in of  the highest peak to emphasize the smoothness of the numerical 
solutions. The time scales of the relevant plots are respectively $\Delta t 
\approx 10^{-1}$ around $t = 9.7$ (top right panel) and $\Delta t \approx 
10^{-4}$ around $t = 9.7$ (bottom right panel).}

% \label{Fig:rho_sigma}
\end{figure}

From the modified Klein-Gordon equation \eqref{eom}, at the turning point,
we have:
\begin{equation}
\dot{\varphi} = - \frac{V_{, \phi}}{(1-g)}.
\label{sign_dotphi}
\end{equation}
Differentiating $V(\phi)$, one can show that the sign of 
$V_{,\phi}$ is opposite to that of $v$, defined through
\begin{equation}
v \equiv 1 - \bV \ex^{(1+\bV)\sqrt{\frac{2}{q}} \phi},
\label{defv}
\end{equation}
as can be seen from
\begin{equation}
V_{,\phi} = -2 V_0 \sqrt{\frac{2}{q}} 
\[ 1+\ex^{\sqrt {\frac{2}{q}} (1 + \bV) \phi}\]^{-2} v.
\end{equation}
The definition \eqref{defv} implies that $v$ is positive (respectively
negative) if $\phi$ is less (resp. greater) than $\phi_\mathrm{lim}$
defined by
\begin{equation}
\phi_{\mathrm{lim}} = -\sqrt{\frac{q}{2}} \frac{\ln \( \bV\)}{1+\bV}  .
\label{def_phi_lim}
\end{equation}
From Eq.~\eqref{sign_dotphi}, $\mathrm{sign}(\dot{\varphi}) =
\mathrm{sign}(v)$, since outside the ghost condensate phase, $(1-g) >
0$. We therefore conclude that, at the turning point,
$\dot{\varphi}(t_{\mathrm{T}})$ is positive (resp. negative) if $\phi
(t_{\mathrm{T}})$ is less (resp. greater) than $\phi_{\mathrm{lim}}$.

Evaluating Eq.~\eqref{def_poly} in $y_{\mathrm{lim}} =
y(\phi_{\mathrm{lim}})= \bV^{-1/(1+\bV)}$, we have
\begin{equation}
f(y_{\mathrm{lim}}) =  \bV^{\frac{1}{1+\bV}} + \bV^{-\frac{\bV}{1+\bV}}
+ \frac{2}{3} \frac{V_0}{H^2 - \displaystyle\frac{\sigma^2}{6}},
\end{equation}
so that, for the chosen parameters, we obtain
$$
\bV^{\frac{1}{1+\bV}} + \bV^{-\frac{\bV}{\bV+1}} \approx 1.5 \sim
\mathcal{O}(1).
$$
As can be read from the graph, $H^2$ is of order $10^{-11} - 10^{-12}$
around the turning point (marked as $t_{\mathrm{T}}$ in
Fig.~\ref{Fig:rhos}), whereas $\rho_{\sigma} \approx 10^{-10}$, so we
can neglect the former with respect to the latter and assume $H^2 <
\sigma^2/6$ for an order of magnitude estimate.
We then get 
$$
\frac{2}{3} \frac{V_0}{H^2 - \displaystyle\frac{\sigma^2}{6}}\approx
-\frac{10^{-7}}{10^{-10}} \approx - 10^{3},
$$
and therefore that $f(\phi_\mathrm{lim})<0$. This means, according to
our previous considerations, that Eq.~\eqref{def_poly} admits two roots
$\phi^{(1)}_{\mathrm{T}} \in \big[ \phi_{\mathrm{lim}},  \infty\big[$ and
$\phi^{(2)}_{\mathrm{T}} \in \big]- \infty ,\phi_{\mathrm{lim}}\big]$, only one
of which satisfying the necessary condition on the sign in
$\varphi_\mathrm{T}$ to be a maximum or a minimum.

One should notice that the relative importance of the anisotropy with
respect to the average Hubble parameter is precisely what permits
$f(y)$ to become negative somewhere and hence to have a root, i.e., to
lead to the existence of a turning point. This estimate is in agreement with our 
previous remark that a longer contraction phase is related to scenarios with 
multiple bounce stage: as can be seen on Fig. ~\ref{Fig:rho_sigma}, the single 
bounce case exhibits a relatively small contribution of anisotropy compared with 
the two bounce case. Longer contraction phases
permit the buildup of larger anisotropies, which in turn eases the
condition $\rho_\sigma > 3 H^2$ necessary for the appearance of
a turning point.

\section{Conclusions} \label{Sec:Concl}

Classical non singular bouncing cosmology as a paradigm faces many
problems \cite{Battefeld:2014uga} that need be addressed before any
realistic model can be constructed and seriously compared with the
available data \cite{Ade:2015lrj}. Among the challenges lies the
question of the shear, whose behavior during a contraction phase
endangers any model of a BKL instability irremediably pushing the
dynamics towards a singularity. To date, there is no other means to
cure this potential plague but to invoke a long-enough ekpyrotic phase.
This must be followed by the actual bounce in order to connect the
resulting universe to ours, currently expanding. It appears the most
economical way to do so is to invoke a scalar field $\phi$ whose
potential $V(\phi)$ can drive an ekpyrotic epoch while a non-standard
kinetic term $g(\phi)$ can yield a ghost-condensate phase sufficient to
initiate a null energy condition violation from which a bounce can
result. Such a simple model has already been discussed and analyzed in
\cite{Cai:2012va}, and the question of the evolution of the shear
during the bounce transition was addressed in \cite{Cai:2013vm}.

In this work, we returned to the model of \cite{Cai:2013vm}, and found
the dynamics potentially much richer than previously thought. In
particular, assuming the same underlying microscopic parameters, we
numerically found and presented four different scenarios depending only
on the choice of initial conditions. These are: a singular solution,
following a long contraction phase which increases the anisotropy despite
the presence of an ekpyrotic potential and failing to bounce because of a
too fast ghost condensate phase; a single-bounce solution, already
encountered in the existing literature, in which the universes contracts,
passes through a minimum scale factor and expands again isotropically;
two and three bounce solutions, in which the universe shows many turning
points and consequently passes more than once though the top of the
kinetic coefficient $g(\phi)$ and the potential $V(\phi)$.

As it turns out, the failed bounces are not in fact a limiting situation
of a single bounce case, but rather they are multiple-bounce cases for
which the last turning point yielded a ghost-condensate phase whose
duration was not long enough to actually bounce. Thus, the slope of the
scale factor changes through this phase, with the Hubble scale increasing
in much the same way as during a bounce, except that it never reaches
positive values. The conditions right after this phase are such as to
throw the universe into the singularity.

All but the singular scenarios lead to a final isotropic expansion which
render them indistinguishable from the background point of view if
confronted with observations. However, one should expect severe changes
in the primordial power spectrum, in particular in that the different
regimes we obtained may not only spoil the scale invariance of long
wavelength modes that might have been produced in the early stages, but
also in that it could imprint a privileged direction in this spectrum due
to the fact that the shear is not negligible during many phases of the
evolution. We even found cases for which the turning point, and hence the
very existence of a bounce, was demanding the shear to dominate at some
stage!

There are many potentially observable consequences such a rich
background dynamics may lead to, that should be derived and
subsequently either confronted with the data or constrained by them. In
particular, since the shear is not necessarily negligible at all times,
and because there is a long and crucial contraction phase, vector modes
can be produced which should be limited in order not to spoil the
bounce and the following isotropic expansion. Besides, couplings
between the scalar, vector and tensor modes could trigger new imprints
and correlations \cite{Pereira:2007yy}, whose exact properties and
characteristic features should be provided by a more complete and
thorough analysis. As we have seen, the background dynamics seems very
sensitive (chaotic?) to the initial conditions on the scalar field, and
it may well be that this sensitivity also transfers to the
perturbations. The negative side of this fact is that the models are
probably not as generic as one would have wanted them to be, but this
also means a positive side, namely that some a priori unwanted
consequences may induce very easily identifiable effects, either in the
perturbation spectra (e.g., specific correlations between scalars or
tensors going beyond the consistency relation) or in higher order
functions (non gaussianities) \cite{Karciauskas:2008bc}. Finally, we
should like to mention that because this category of models can lead to
a potentially observable privileged direction in the expanding
universe, this could also induce large scale anomalies that should be
compared with those present in the existing or future observations, for
instance in the cosmic microwave background data
\cite{Ade:2013nlj,Schwarz:2015cma}.

\acknowledgments

PP and SDPV would like to thank the Labex Institut Lagrange de Paris
(reference ANR-10-LABX-63) part of the Idex SUPER, within which this work
has been partly done. SDPV acknowledges the financial support from BELSPO
non-EU postdoctoral fellowship and from the CNPq-Brazil PCI/MCTI/CBPF
program. APB would like to thank CNPq-Brazil for the financial support through 
the program ``Ci\^{e}ncias sem Fronteiras'' (grant no. 233560/2014-9) and the 
Institut d'Astrophysique de Paris for  
hosting her during the development of this work. 

\bibliographystyle{apsrev}

\end{document}